\documentclass[11pt,article,aps,prd,groupedaddress,preprint,eqsecnum, nofootinbib]{revtex4}
\usepackage{graphicx,epsf,amssymb,amsbsy,amsfonts,amssymb,amsmath,setspace}

\def\be{\begin{equation}}
\def\ee{\end{equation}}

\def\m{\mu}\def\n{\nu}

\def\bg{\bar{g}}
\def\beq{\begin{eqnarray}}\def\eeq{\end{eqnarray}}
\def\ba#1\ea{\begin{align}#1\end{align}}
\def\bg#1\eg{\begin{gather}#1\end{gather}}
\def\bm#1\em{\begin{multline}#1\end{multline}}
\def\bmd#1\emd{\begin{multlined}#1\end{multlined}}

\def\m{\mu}
\def\n{\nu}

\def\({\left(}
\def\){\right)}
\def\[{\left[}
\def\]{\right]}

\def\m{\mu}
\def\n{\noindent}

\begin{document}
\hfuzz 12pt

\title{RG Flow and Thermodynamics of Causal Horizons in Higher-Derivative AdS Gravity}
\author{Shamik Banerjee}
\affiliation{Kavli Institute for the Physics and Mathematics of the Universe, The University of Tokyo,\\
5-1-5 Kashiwa-no-Ha, Kashiwa City, Chiba 277-8568, Japan }

\email{banerjeeshamik.phy@gmail.com}

\author{Arpan Bhattacharyya}

\affiliation{Centre for High Energy Physics, Indian Institute of Science, 
C.V. Raman Avenue, Bangalore 560012, Karnataka, India}

\email{bhattacharyya.arpan@yahoo.com}  

\begin{abstract}
In arXiv:1508.01343 [hep-th], one of the authors proposed that in AdS/CFT the gravity dual of the boundary $c$-theorem is the second law of thermodynamics satisfied by causal horizons in AdS and this was verified for Einstein gravity in the bulk. In this paper we verify this for higher derivative theories. We pick up theories for which an entropy expression satisfying the second law exists and show that the entropy density evaluated on the causal horizon in a RG flow geometry is a holographic c-function. We also prove that given a theory of gravity described by a local covariant action in the bulk a sufficient condition to ensure holographic c-theorem is that the second law of causal horizon thermodynamics be satisfied by the theory. This allows us to explicitly construct holographic c-function in a theory where there is curvature coupling between gravity and matter and standard null energy condition cannot be defined although second law is known to hold. Based on the duality between c-theorem and the second law of causal horizon thermodynamics proposed in arXiv:1508.01343 [hep-th] and the supporting calculations of this paper we conjecture that every Unitary higher derivative theory of gravity in AdS satisfies the second law of causal horizon thermodynamics. If this is not true then c-theorem will be violated in a unitary Lorentz invariant field theory. 
\end{abstract}

\preprint{IPMU15-0169}
\maketitle
\tableofcontents

\section{Introduction}
Renormalization group (RG) flow describes the change in the effective behaviour of a quantum field theory as the distance or energy scale is changed. In the Wilsonian scheme one starts from a bare theory in the ultraviolet (UV) and integrates out high energy degrees of freedom. The effect of integrating out the high energy degrees of freedom is to redefine or renormalize the bare coupling constants appearing in the Lagrangian of the UV theory. This redefinition is mathematically described as a first order flow equation in the space of coupling constants. Conformal field theories (CFT) appear as fixed points of this flow. Now given a CFT (UV-CFT) one can perturb this by adding a relevant operator and flow to a new CFT (IR-CFT) in the infrared (IR). For two dimensional unitary and Lorentz invariant quantum field theories it has been shown in \cite{Zam} that one can define a positive monotonically decreasing function of the coupling constants along the flow whose values at the fixed points coincide with the central charges of the respective two dimensional CFTs. This positive monotonically decreasing function of the coupling constants is known as a $c$-function. $c$-theorem was generalized and proved for higher dimensional field theories in \cite{Cappelli:1990yc,Cardy:1988cwa, ctheorems, Myers:2010xs}.

RG flow can also be studied via AdS/CFT correspondence. In this set up one usually starts with a bulk theory which has several pure AdS critical points. The bulk theory is usually a classical theory of gravity coupled to matter fields. RG flow is represented by a domain-wall like solution of the coupled gravity-matter system which interpolates between two AdS critical points \cite{Akhmedov:1998vf, Freedman:1999gp,cthor,ctheorems, Myers:2010xs,deBoer:1999xf}. \footnote{There is a different approach to studying the holographic RG flow using the holographic entanglement entropy \cite{Ryu:2006bv,chm,LM}. We will refer to \cite{Liu,Myers:2012ed} for further details on this. }The holographic version of the $c$-theorem is proved by showing the existence of a function which decreases monotonically from the UV-AdS to the IR-AdS as a result of the field equations and gives the central charges when evaluated at the end points of the flow. In all these constructions null energy condition on the matter stress tensor plays an important role. 

The story of the holographic $c$-theorem is more interesting when one considers higher derivative corrections to Einstein gravity in the bulk. This was considered in great detail in \cite{ctheorems}. Higher derivative corrections are ubiquitous in low energy effective theories coming from string theory and  they are also interesting in their own right as toy models where one can study various thorny issues related to black hole thermodynamics. In an arbitrary theory of gravity described by a local covariant action there is in general be no clear separation between gravity and matter sector. So matter stress tensor cannot be uniquely defined and standard null energy condition has no clear role to play. This is one of the main obstacles of proving holographic $c$-theorem in a general theory as was emphasized in \cite{ctheorems}.  The other point is that even if there is a higher derivative theory of gravity coupled to matter where null energy condition can be imposed unambiguously, sometimes it is not enough to prove holographic $c$-theorem and validity of the $c$-theorem seems to require additional constraints on the pure gravity sector of the theory \cite{Liu:2010xc}. This could be a genuine physical constraint on the bulk theory like unitarity but its status is always not so clear \cite{Liu:2010xc}. So an important problem is to spell out the constraint which needs to be imposed on a theory of gravity so that holographic $c$-theorem holds.

In \cite{sb}, a different approach towards holographic RG flow and holographic $c$-theorem was proposed. The basic observation of \cite{sb} was that causal horizons in pure Poincare AdS are Killing horizons generated by the dilatation vector. Dilatation symmetry is broken along the RG flow and as a result causal horizons in a RG flow geometry are no longer globally Killing. They become Killing (stationary) in the asymptotic UV and IR region of the RG flow geometry, where scaling is recovered. In \cite{sb} it was shown that the Bekenstein-Hawking entropy (density) of the non-Killing causal horizon interpolating between the UV and the IR of the RG flow geometry is a holographic $c$-function. The monotonicity of the $c$-function followed from the second law of thermodynamics obeyed by causal horizons in AdS \cite{Jacobson:2003wv}. Based on these observations it was proposed in \cite{sb} that the boundary RG flow is dual to the thermodynamics of causal horizons in AdS and the gravity dual of the $c$-theorem is the second law of causal horizon thermodynamics. 

In this paper we verify this for higher derivative theories of gravity which admit pure $AdS$ as a solution. At first we test it for higher derivative theories for which an entropy expression exists which satisfies at least the linearized version of the second law. It has been shown recently \cite{Sarkar,Sarkar1,Sarkar2,Sarkar3} \footnote{The entropy expression which obeys the linearized second law are derived from the holographic entanglement entropy functionals obtained in \cite{bss, Fursaev, Dong, Camps}.} that a general curvature-squared theory of gravity satisfies the linearized second law. We verify that in these cases the entropy (density) evaluated on the spatial slices of the dynamical causal horizon is a holographic $c$-function and the gravity dual of the boundary $c$-theorem is the linearized second law. This restricts us to RG flows where the difference between the UV and the IR central charges is small, but there is no restriction on the magnitude of the higher-derivative coupling. In all these cases the bulk gravity theory is minimally coupled to the matter sector which cause the RG-flow and the matter stress tensor is restricted to satisfy the null-energy condition. We would like to emphasize that this does \textit{not} mean that we restrict to gravity-matter theories which satisfy the null convergence condition \footnote{Null convergence condition means that $R_{AB}k^A k^B \ge 0$ for any future directed null vector $k$, where $R_{AB}$ is the Ricci tensor. If this condition is satisfied then Raychaudhuri equation applied to a hypersurface orthogonal congruence of null geodesics implies that the expansion $\theta_D$ is monotonically decreasing along the generators of the congruence. When Einstein gravity is minimally coupled to the matter field the null convergence condition is guaranteed if the matter stress tensor satisfies the null energy condition. However this is not the case if there are higher derivative couplings. Even if the matter field is minimally coupled and the matter stress tensor satisfies the null energy condition, this does not guarantee the null convergence condition due to the presence of the higher derivative terms. So focusing does not happen and the proof of the second law cannot follow the same logic  as in the Einstein gravity case.} which is generically violated in all higher-derivative theories of gravity including the ones we are studying. 

Our next example is $F(R)$ gravity where $R$ is the Ricci scalar. In this case the entropy expression satisfying the second law to all orders is known \cite{jkm} and we construct the holographic $c$-function using this. This is an example of a theory which has fourth order equation of motion, but the bulk theory is still unitary \cite{Stelle:1977ry, Nakasone:2009vt}.

We also study the effect of non-minimal coupling to the matter sector.  As an example, we treat the case of Einstein gravity plus scalar matter where in addition to the standard minimal coupling there is also a curvature coupling of the form $R\phi^2$, where $R$ is the Ricci scalar. In this case there is no clean separation between matter and gravity sector and as a result standard null energy condition cannot be defined. In such a theory the second law is known to hold, albeit with an expression which is not the area of a space-like slice of the horizon \cite{Ford}. Using this expression we prove holographic $c$-theorem in a situation where the null-energy condition cannot be defined. As a generalization of this we show that a sufficient condition to ensure holographic $c$-theorem in a theory of gravity described by a local covariant action in AdS is that the second law of causal  horizon thermodynamics be satisfied by the theory.

\section{RG flow and Causal horizon thermodynamics}

\begin{figure}[htbp]
\begin{center}
  \includegraphics[width=15cm]{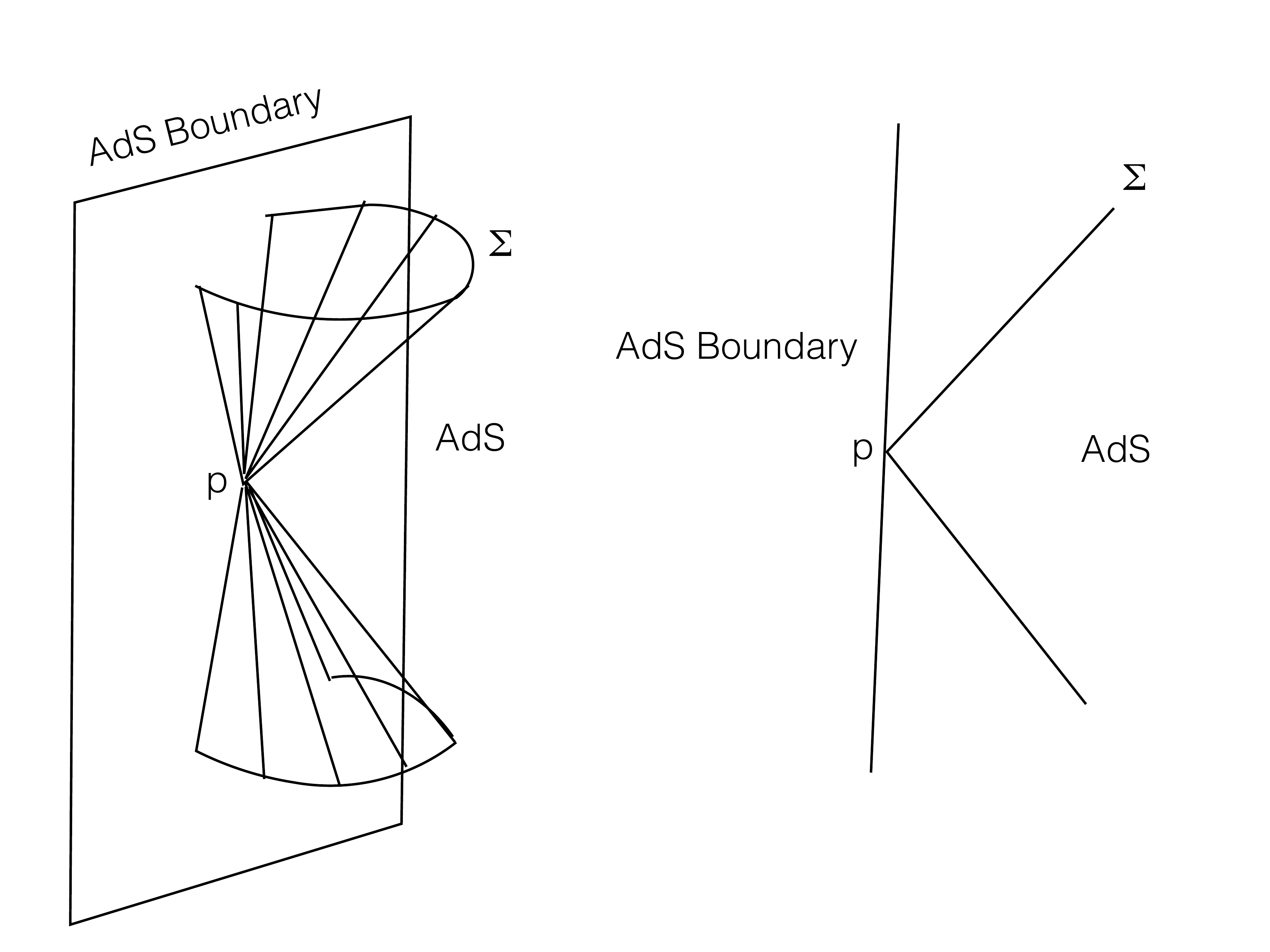}
\end{center}
\caption{Causal horizon $\Sigma$  of the boundary point p in pure AdS. This is just the bulk light-cone of the point p. Poincare time runs vertically upward in the diagram.}
\end{figure}

In this section we will briefly review the contents of \cite{sb} and discuss some notations.
Let us consider $AdS_{d+1}$ written in Poincare coordinates,
\be
ds^2=\frac{L_{AdS}^2}{z^2}(-dt^2+ \sum_{i=1}^{d-1}(dx^i)^2+ dz^2)
\ee
where $L_{AdS}$ is the $AdS$ radius. It will be useful to introduce the spherical polar coordinates such that, 

\be
\sum_{i=1}^{d-1} (dx^i)^2 =  dr^2+ r^2d\Omega_{d-2}^2
\ee

where, $d\Omega_{d-2}^2$ is the metric of a unit  $(d-2)$-sphere. In these coordinates, $AdS$-metric is,

\be
ds^2=\frac{L_{AdS}^2}{z^2}(-dt^2+ dr^2+ r^2d\Omega_{d-2}^2+ dz^2).
\ee

This metric has a scaling isometry, $(t,r,z)\rightarrow(\lambda t,\lambda r, \lambda z)$, generated by the dilatation vector,

\be
D=\frac{t}{2}\frac{\partial}{\partial t}+\frac{r}{2} \frac{\partial}{\partial r}+\frac{z}{2} \frac{\partial}{\partial z}
\ee

$D$ is null on the hypersurface $\Sigma$ given by,

\be
-t^2 + r^2 + z^2 =0 
\ee

$\Sigma$ is a null hypersurface which is left invariant by the flow of $D$ and so $\Sigma$ is a Killing horizon generated by $D$.  It is also easy to see that $\Sigma$ is the boundary of the causal past (future) of the point, $t=x^i=z=0$, in $AdS$.  So $\Sigma$ is a causal horizon and in pure $AdS$ it is a Killing horizon generated by the dilatation vector, $D$. 

Let us now deform the boundary field theory by adding a relevant operator which is dual to a scalar field of negative mass squared in $AdS$. This induces RG-flow. 
For simplicity we will take the backreacted geometry to be,

\be\label{domain}
ds^2=\frac{L_{AdS}^2}{z^2}\Big(-dt^2+ dr^2+ r^2d\Omega_{d-2}^2+ \frac{dz^2}{f(z)}\Big)
\ee

where $f(z)\rightarrow 1$ as $z\rightarrow 0$ and $f(z)\rightarrow \frac{L_{AdS}^2}{L_{IR}^2}$ as $z\rightarrow \infty$. Here $L_{IR}$ is the radius of the locally $AdS$ region which develops near the Poincare horizon, $z\rightarrow\infty$. This metric has no global scaling isometry because dilatation is broken along the RG-flow.  Scaling is recovered only asymptotically as $z\rightarrow 0$ or $z\rightarrow \infty$.

Let us now write the metric in the form,

\be\label{remetric}
ds^2=\frac{L_{AdS}^2}{z^2}(-dt^2+ dr^2+ r^2d\Omega_{2}^2+ d\bar z^2)
\ee

where we have defined,

\be\label{barz}
\frac{\bar z}{z}=\frac{1}{z}\int_{0}^{z} \frac{dz'}{\sqrt{f(z')}}=\int_{0}^{1}\frac{d\alpha}{\sqrt{f(\alpha z)}}
\ee

In (\ref{remetric}), $z$ should be treated as a function of $\bar z$ given implicitly by the relation (\ref{barz}). Throughout the paper prime over $z$ will denote differentiation with respect to $\bar z$.

The causal horizon of the point, $t=x^i=z=0$, in the back-reacted geometry can now be written as,

\be\label{dcausal}
-t^2 + r^2 + \bar z^2 =0
\ee

\begin{figure}[htbp]
\begin{center}
  \includegraphics[width=13cm]{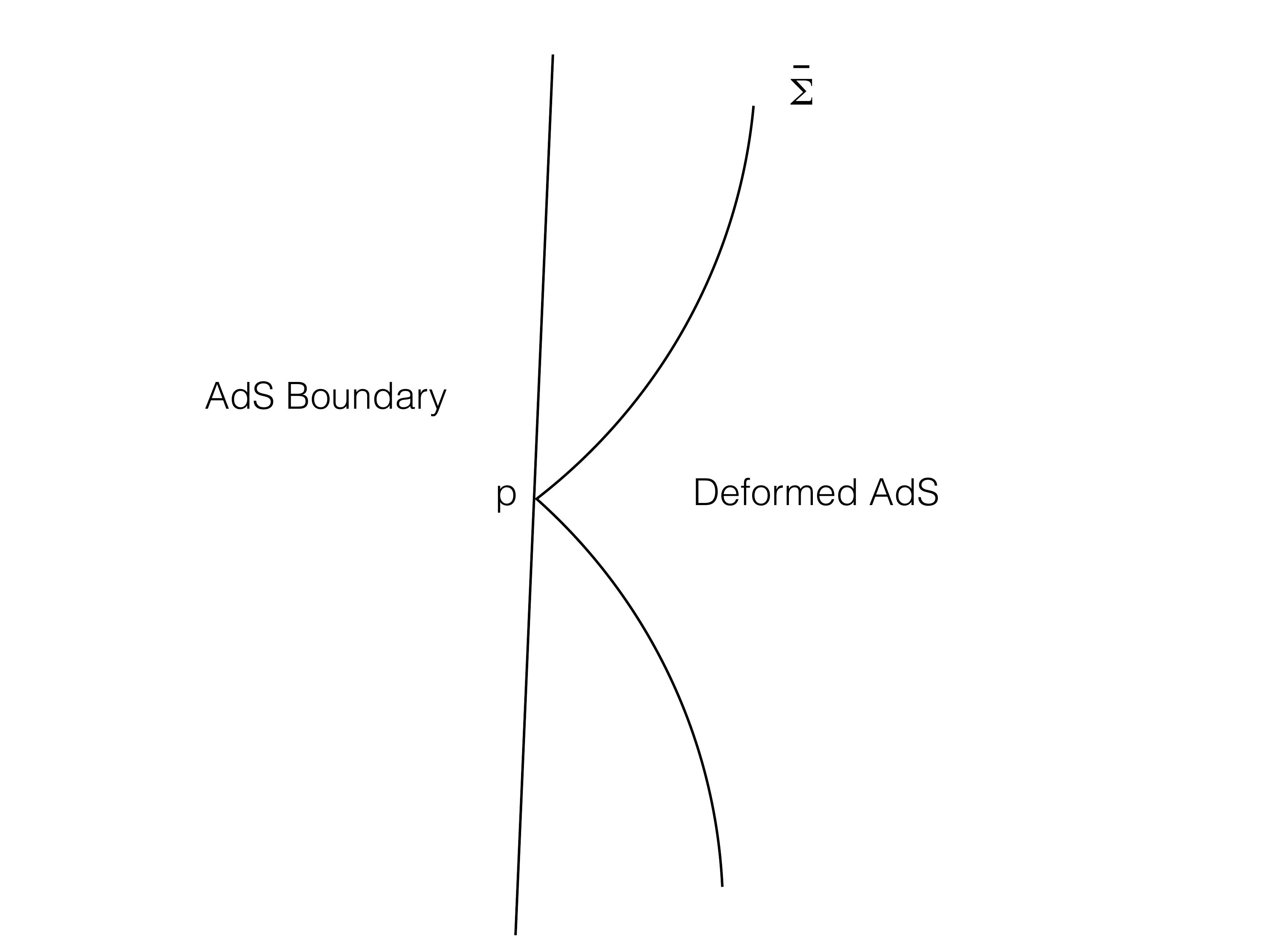}
\end{center}
\caption{Deformed causal horizon $\bar\Sigma$  of the boundary point p in deformed AdS. This is just the deformed bulk light-cone of the point p. Poincare time runs vertically upward in the diagram.}
\end{figure}

Let us denote this by $\bar\Sigma$. $\bar\Sigma$ is again a null-hypersurface and we can take as its null generator the vector field $\bar D$ given by, \cite{sb}

\be
\bar D=\frac{t}{2}\frac{\partial}{\partial t}+\frac{r}{2} \frac{\partial}{\partial r}+\frac{\bar z}{2} \frac{\partial}{\partial\bar z}
\ee 

$\bar D$ is null on $\bar\Sigma$. $\bar D$ is not a Killing vector. It coincides with $D$ only asymptotically where scaling is recovered.  As a result the backreacted causal horizon $\bar\Sigma$ is \textbf{no} longer a Killing horizon. It is dynamical in the sense that it has non-zero expansion and shear. 

In \cite{sb} it was proposed that the boundary RG-flow is dual to the (thermo) dynamics of causal horizons caused by the breaking of the scaling isometry in the bulk. In particular it was shown in \cite{sb} that if the bulk theory is described by Einstein-Hilbert action minimally coupled to a scalar field obeying the null-energy condition then the gravity dual of the boundary $c$-theorem is the second law of thermodynamics obeyed by causal horizons \cite{Gibbons:1977mu,Jacobson:2003wv}. The holographic $c$-function in this framework was given by the Bekenstein-Hawking entropy density of the dynamical causal horizon (\ref{dcausal}). \textit{This directly relates the second law for causal horizons in $AdS$ to the unitarity of the boundary theory in the form of $c$-theorem}.


\section{Holographic $c$ - function in higher derivative gravity}
\subsection{Gauss-Bonnet gravity}




We first consider the  Einstein-Gauss-Bonnet theory in $AdS_5$ given by the action,
\be
I_{GB}=-\frac{1}{2\ell_{p}^3}\int d^{5}x \sqrt{g}\Big(R+\frac{12}{L^2}+\frac{\lambda\, L^2}{2}(R_{\alpha\beta\gamma\delta}R^{\alpha\beta\gamma\delta}-4R_{\alpha\beta}R^{\alpha\beta}+R^2)\Big).
\ee

If we minimally couple it to scalar matter which induces the RG-flow, then the equation of motion for this theory is given by,
\be \label{met3}
R_{\alpha\beta}-\frac{1}{2} R g_{\alpha\beta}-\frac{6}{L^2} g_{\alpha\beta}-\frac{\lambda\,L^2}{2}\, H_{\alpha\beta}=T_{\alpha\beta}
\ee
where,
\begin{align}
\begin{split}
H_{\alpha\beta}= &\,\,4R_{\alpha}^{\delta}R_{\delta\beta}-4 R^{\delta\sigma}R_{\delta\alpha\beta\sigma}
-2 RR_{\alpha\beta}-2 R_{\alpha\delta\sigma\mu}R_{\beta}{}^{\delta\sigma\mu}\\&+\frac{1}{2}g_{\alpha\beta}(R_{\alpha\beta\gamma\delta}R^{\alpha\beta\gamma\delta}-4R_{\alpha\beta}R^{\alpha\beta}+R^2).
\end{split}
\end{align}
 and $T_{\alpha\beta}$ is the matter stress tensor.  We have assumed that the matter stress-tensor obeys the null-energy condition. So if we contract this with the null vector $\bar D$ 
we will get,
\be \label{met2}
R_{\alpha\beta}\bar D^{\alpha}\bar D^{\beta}-\frac{\lambda\,L^2}{2}\,H_{\alpha\beta}\bar D^{\alpha}\bar D^{\beta}\geq 0.
\ee
Evaluating (\ref{met2}) on (\ref{remetric}) we get the following condition,
\be \label{met4}
\frac{3 \bar z^2 f'(z) (1-2 f_{\infty} \lambda  f(z))}{8\,\, z}\geq 0.
\ee

where $f_{\infty}= \frac{L^2}{L_{AdS}^2}$. We will analyze this condition later.

Let us now turn our attention to the entropy functional which is the Jacobson-Myers (JM) one in this case \cite{jkm, Hung, bss}
\be
S_{JM}=\frac{2\pi}{\ell_{p}^3}\int d^{3} x \sqrt{h}\Big(1+\lambda\, L^2\,\mathcal{R}\Big).
\ee 

The integral is taken along a space-like slice of the horizon (\ref{dcausal}) and $\mathcal{R}$ is the intrinsic Ricci scalar of the space-like slice. JM entropy functional satisfies the linearized second law. To evaluate this, it is useful to parametrize the causal horizon (\ref{dcausal}) as,
 
\be\label{parahyp}
t =\bar z \cosh\eta\,, \quad r=\bar z \sinh\eta,\quad \bar z=\bar z
\ee

This parametrization corresponds to a particular space-like slicing of the causal horizon.

With this parametrization the induced metric on the space-like slices of (\ref{dcausal}) becomes, \footnote{This is also the induced metric on the null hypersurface $\bar\Sigma$.}
\be \label{hyp}
ds_{ind}^2= L_{AdS}^2 \Big(\frac{  \bar z }{z}\Big)^2\Big( d\eta^2+ \sinh^2 \eta \ d\Omega_2^2\Big) =  L_{AdS}^2 \Big(\frac{\bar z}{z}\Big)^2 ds_{H^3}^2
\ee

where $ds_{H^3}^2$ is the metric on a  unit hyperboloid.

So,
\be
\mathcal R = -\frac{6}{{L_{AdS}}^2} \Big({\frac{z }{\bar z}}\Big)^2 
 \ee
 
Now following \cite{sb} we take the null generators of the causal horizon as integral curves of $\bar D$ defined as,

\be\label{inte}
\frac{dt}{d\sigma} = t, \ \frac{dr}{d\sigma} = r, \ \frac{d\bar z}{d\sigma} = \bar z, 
\ee
where $\sigma$ is not necessarily an affine parameter. The triple $(t(\sigma),r(\sigma),\bar z(\sigma))$ satisfies the condition,
\be\label{nullcondition}
-t(\sigma)^2 + r(\sigma)^2 + \bar z(\sigma)^2 = 0
\ee

It follows from (\ref{parahyp}) and (\ref{inte}) that $\eta$ is a comoving coordinate, i.e $\eta$ remains constant along the integral curves satisfying (\ref{inte}) and (\ref{nullcondition}) and so all the coordinates parametrizing the unit hyperbolic space are comoving.


Let us now define the density of JM entropy of the causal horizon as,

\be
dS_{JM} = \frac{2\pi}{\ell_{p}^3} d^{3} x \sqrt{h}\Big(1+\lambda\, L^2\,\mathcal{R}\Big) = 2\pi \frac{L_{AdS}^3}{\ell_{p}^3} \Big(\frac{\bar z}{z}\Big)^3 \Big(1- 6\lambda  \frac{L^2}{L_{AdS}^2} \Big(\frac{z}{\bar z}\Big)^2 \Big) \ dV_{H^3}
\ee

Since the volume of the unit hyperboloid is independent of the parameter $\sigma$ of the null generator, second law becomes a statement about the evolution of the finite coefficient along the null generator.  We denote the finite coefficient by $a(\sigma)$, 

\be\label{GBc}
 a(\sigma) = 2\pi \frac{L_{AdS}^3}{\ell_{p}^3} \Big(\frac{\bar z}{z}\Big)^3 \Big(1- 6\lambda  \frac{L^2}{L_{AdS}^2} \Big(\frac{z}{\bar z}\Big)^2 \Big)= 2\pi \frac{L_{AdS}^3}{\ell_{p}^3} \Big(\frac{\bar z}{z}\Big)^3 \Big(1- 6\lambda f_{\infty} \Big(\frac{z}{\bar z}\Big)^2 \Big)
\ee

This gives the correct $a$-central charge at the UV ($\sigma\rightarrow -\infty$ ) and IR ($\sigma\rightarrow \infty$) \footnote{It follows from (\ref{inte}) that as $\sigma\rightarrow -\infty$, $\bar z\rightarrow 0$ and as $\sigma\rightarrow\infty$, $\bar z\rightarrow\infty$. It follows from a simple argument of \cite{sb} that $z\rightarrow 0$ as $\bar z\rightarrow 0$ and $z\rightarrow \infty$ as $\bar z\rightarrow \infty$. So as $\sigma$ runs from $-\infty$ to $+\infty$, the null generator of the causal horizon runs from the UV to the IR.} fixed points. To see this more quantitatively \cite{sb}, we note that as $\sigma\rightarrow -\infty$, $z\rightarrow 0$ and $\frac{\bar z}{z}\rightarrow 1$. Similarly as $\sigma\rightarrow\infty$, $z\rightarrow\infty$ and $\frac{\bar z}{z}\rightarrow \frac{L_{IR}}{L_{AdS}}$ . Substituting this we get the correct values of the UV and the IR central charges, respectively \footnote{In the context of dS/CFT in presence of Gauss-Bonnet correction it will be interesting to investigate the $c$ function following the references of \cite{Nojiri:2001ae}  where the expression for the entropy and the central charge have been worked out.}.

The monotonicity follows from the linearized second law satisfied by the JM entropy function, but we will check this by explicit calculation. Since the JM entropy satisfies the linearized second law, the $c$-function (\ref{GBc}) constructed out of the JM entropy function is monotonic if we confine ourselves to linearized metric fluctuations. By linearized we mean that the metric fluctuations around the UV-AdS is of the same order of magnitude as the matter stress tensor causing the RG flow. This condition allows us to probe holographic RG flows where the difference between the UV and the IR central charges is small. This does \textbf{not} mean that Gauss-Bonnet coupling constant $\lambda$ is small. 

Now,
\be \label{avariation}
\frac{d a(\sigma)}{d\sigma}=3 \left(\frac{\bar z}{z }\right) \left(1-\int_{0}^{1}\frac{d\alpha\, \sqrt{f(z)}}{\sqrt{f(\alpha z)}}\right) \left(\Big(\int_{0}^{1}\frac{d\alpha}{\sqrt{f(\alpha z)}}\Big)^2-2f_{\infty}\lambda \right). 
\ee

We want to show that,

\begin{equation}\label{monotony}
\frac{da(\sigma)}{d\sigma} \le 0
\end{equation}

One can check that this condition cannot be proved in complete generality without imposing an extra condition that $f'(z)\ge 0$. But this does not follow from the null energy condition, 

\be
\frac{3 \bar z^2 f'(z) (1-2 f_{\infty} \lambda  f(z))}{8\,\, z}\geq 0.
\ee
Since $\frac{\bar z^2}{z}$ is positive this condition is equivalent to, 

\be \label{NE}
f'(z) (1-2 f_{\infty} \lambda  f(z))\geq 0.
\ee

The fact that we cannot prove (\ref{monotony}) in complete generality using the condition (\ref{NE}) is just the reflection of the fact that the JM entropy functional satisfies only the linearized second law.  So let us do a linearized calculation. Let us assume that the matter stress tensor is of order $\epsilon$. Since in the fixed point geometry $f(z)$ is a constant, the leading term of $f'(z)$ is of order $\epsilon$ and linearized calculation means we keep only order $\epsilon$ terms in our calculation. So in this approximation, the null-energy condition becomes, 

\be\label{LNE}
f'(z) (1-2 f_{\infty} \lambda )\geq 0.
\ee 

Applying the linearization to (\ref{avariation}) we get,

\be
\frac{d a(\sigma)}{d\sigma}=3 \left(1-\int_{0}^{1}\frac{d\alpha\, \sqrt{f(z)}}{\sqrt{f(\alpha z)}}\right) \left(1 - 2f_{\infty}\lambda \right). 
\ee

Now we can write,

\be
1-\int_{0}^{1}\frac{d\alpha\, \sqrt{f(z)}}{\sqrt{f(\alpha z)}}  = - \frac{1}{2} \int_{0}^{1} \frac{d\alpha}{\sqrt{f(\alpha z)}} \int_{\alpha z}^{z} \frac{d\beta}{\sqrt{f(\beta)}} f'(\beta) 
\ee

Using this we can write, 

\be \label{Savariation}
\frac{d a(\sigma)}{d\sigma}= -\frac{3}{2} \int_{0}^{1} \frac{d\alpha}{\sqrt{f(\alpha z)}} \int_{\alpha z}^{z} \frac{d\beta}{\sqrt{f(\beta)}} f'(\beta) 
\left(1 - 2f_{\infty}\lambda \right). 
\ee

The expression on the RHS of (\ref{Savariation}) is manifestly negative once we apply the linearized null-energy condition (\ref{LNE}). So, 

\be
\frac{da(\sigma)}{d\sigma} \le 0 
\ee

to linearized order. So the $c$-function constructed from JM entropy functional decreases monotonically as a consequence of the linearized second law for causal horizons. This is the best we can hope from the JM entropy functional. 

\subsection{General curvature squared gravity}




We now consider the general curvature squared theory in $AdS_5$ given by the action,
\be
I_{R^2}=-\frac{1}{2\ell_{p}^3}\int d^{5}x \sqrt{g}\Big(R+\frac{12}{L^2}+\frac{L^2}{2}\Big(\lambda_{1}R_{\alpha\beta\gamma\delta}R^{\alpha\beta\gamma\delta}+\lambda_{2} R_{\alpha\beta}R^{\alpha\beta}+\lambda_{3} R^2\Big)\Big).
\ee

We minimally couple it to scalar matter which induces the RG-flow.
An entropy functional which satisfies the second law to all orders in this theory is not known. However, an entropy functional which satisfies the linearized second law has been written down recently in \cite{bss, Fursaev, Dong, Camps}. 

In order to write it down we define another auxiliary vector $l^{a}$ such that $\bar D_{a}l^{a}=-1$ and $l_{a}l^{a}=0$ on the horizon (\ref{dcausal}). Also $l^{a}$ is orthogonal to the spacelike slice of the horizon. So,
\be
l= \frac{z^2}{\bar z^2} \Big( t\frac{\partial}{\partial t}+ r\frac{\partial }{\partial r}- \bar z\frac{\partial}{\partial \bar z} \Big)
\ee
 With this the entropy functional for this theory can be written as \cite{bss, Fursaev, Dong, Camps} ,
 \be \label{entr2}
S =\frac{2\pi}{\ell_{p}^{3}}\int d^3 x\sqrt{h}\Big(1+ L^2 \Big(\lambda_{3}R-4\lambda_{2} R_{\alpha\beta} \bar D^{\alpha}l^{\beta}+\lambda_{1}R_{\alpha\beta\gamma\delta}\bar D^{\alpha} l^{\beta} \bar D^{\gamma} l^{\delta}+\frac{1}{6} (4 \lambda_{1}+3 \lambda_{2})\theta_{\bar D}\theta_{l}\Big)\Big).
\ee

where $\theta_{\bar D}$ and $\theta_{l}$ are respectively the two expansion along $\bar D$ and $l$ given by,
\be
\theta_{\bar D}=\frac{3}{2}-\frac{3 \bar z  z'}{2 z}
\ee
 and 
\be
\theta_{l}=\frac{3 z \left(\bar z z'+z\right)}{\bar z^2 L_{AdS}^2}.
\ee

In general the shear term also enter in the entropy functional, but in our case it vanishes.
 
We get the $c$-function after evaluating (\ref{entr2}) on (\ref{hyp}),
\be \label{c2}
a(\sigma)=2\pi\frac{L_{AdS}^3}{\ell_{p}^3}\Big(\frac{\bar z}{z}\Big)^3\Big(\frac{2 f_{\infty} \bar z^2 z a_{1} z''+\bar z^2 \big(4-5 f_{\infty} a_{1} z'^2\big)+3 f_{\infty} (4 \lambda_{1}+3 \lambda_{2}) z^2}{4 \bar z^2}\Big).
\ee

where $f_{\infty} = \frac{L^2}{L_{AdS}^2}$ and $a_{1}=(4 \lambda_{1}+5 \lambda_{2}+16 \lambda_{3})$.
It gives correct central charges at the fixed points. 
\be
a=2\pi\frac{L_{AdS}^3}{\ell_{p}^3}\Big(1-2 f_{\infty}(\lambda_{1}+2(\lambda_{2}+5\lambda_{3}))\Big).
\ee
It has been shown in \cite{Sarkar1,Sarkar2,Sarkar3} that (\ref{entr2})  obeys linearized second law and as a consequence it follows that $a(\sigma)$ is monotonic.

\section{$c$ function for $F(R)$ gravity}
In this section we will consider the example of $F(R)$ gravity which is an example of higher curvature theory for which the entropy expression satisfying the second law to all orders is known \cite{jkm}. This is also a theory whose the equation of motion is fourth order, but it still gives rise to a \textit{unitary} bulk theory \cite{Stelle:1977ry, Nakasone:2009vt,Biswas:2011ar,Talaganis:2014ida}. We will construct the corresponding $c$ function in this section.  

The action in 5 dimensions is,
\be
I=-\frac{1}{2\ell_{p}^3}\int d^{5}x\sqrt{g}\Big[R+F(R)\Big].
\ee

where $R$ is the Ricci scalar. 

This action is minimally coupled to scalar field whose energy-momentum tensor satisfies the null energy condition and this induces the holographic RG flow. The corresponding entropy functional which satisfies the second law is given by \cite{jkm},
\be
S =\frac{2\pi}{\ell_{p}^3}\int d^{3}x\sqrt{h}\Big[1+F'(R)\Big]
\ee
 where the prime denotes derivative with respect to the Ricci scalar $R$. The integral is over the space like slice (\ref{parahyp}) of the causal horizon (\ref{dcausal}). Since the Ricci scalar for the domain wall geometry (\ref{domain}) depends only on $z$ (or $\bar z$), we can repeat the same arguments as in the previous sections and immediately write down the expression of the $c$-function as,
\be\label{aF}
a(\sigma)=\frac{2\pi\, L_{AdS}^3}{\ell_{p}^{3}}\Big(\frac{\bar z}{z}\Big)^3\Big(1+F'(R(\sigma))\Big).
\ee

where

\be
R(\sigma)=\frac{4 \Big( z(\sigma) f'(z(\sigma))-5 f(z(\sigma))\Big)}{L_{\text{AdS}}^2}
\ee

and $f' = \frac{df}{dz}$. In this expression $z(\sigma)$ is given implicitly by the equation following from (\ref{barz}),

\be
\frac{\bar z(\sigma)}{z(\sigma)} = \int_{0}^{1} \frac{d\alpha}{\sqrt{f(\alpha z(\sigma))}}
\ee 

and $\bar z(\sigma)$ satisfies (\ref{inte}). So $R(\sigma)$ is the Ricci scalar of the domain wall geometry evaluated on the null generator of the causal horizon. The monotonicity ($c'(\sigma)\le 0$) of this $c$ function now follows from the proof of the second law in \cite{jkm}.

Let us now make a few comments on the formula (\ref{aF}). The proof of the second law in \cite{jkm} in case of $F(R)$ gravity relied on the crucial assumption, $1+ F'(R) > 0$, besides null energy condition on matter stress tensor. We can see that in the context of holographic RG flow this positivity condition gets translated into the \textbf{unitarity} of the dual boundary theory. If the positivity condition is violated then the $c$-function (\ref{aF}) can be negative which signals non-unitarity. In our case it is the \textit{entropy density} of the dynamical causal horizon (\ref{dcausal}) which plays the role of holographic $c$-function and so the positivity is essential for its interpretation as the number of degrees of freedom of the dual field theory along the RG flow. So we can see that in the case of causal horizons in AdS proof of the second law and unitarity are very closely related. 

We would also like to make a comparison between our formula for the $c$-function (\ref{aF}) and the one proposed in \cite{Liu:2010xc}. The authors of \cite{Liu:2010xc} proposed a formula for the $c$-function in $F(R)$ gravity whose monotonicity relied on the extra assumption which in our notation becomes $F'''(R)\ge 0$. Here prime denotes derivative with respect to $R$. This is a condition on the gravitational sector of the theory. In our case this extra assumption does not play any role. The monotonicity of the $c$-function (\ref{aF}) follows directly from the second law and the second law holds under the assumption of the null energy condition of the matter stress tensor and $1+ F'(R) > 0$. So in this sense our $c$-function is more universal than that proposed in \cite{Liu:2010xc}.

\section{Holographic $c$-theorem in the absence of null energy condition}

Let us consider the following action in 5 dimensions,
\be\label{nmc}
I=-\frac{1}{2\ell_{p}^{3}}\int d^{5}x\sqrt{g}\Big[(1+\zeta \phi^2)R+\frac{12}{L^2}V(\phi)-\frac{1}{2}g^{\mu\nu}\partial_{\mu}\phi\partial_{\nu}\phi\Big].
\ee

where $\phi$ is a scalar field which triggers the RG flow. The importance of more general coupling between gravity and matter was emphasized in \cite{ctheorems}. We assume that this action yields several critical points with constant values of the scalar field and the bulk geometry being pure $AdS_5$. We have a non-trivial curvature coupling, $\zeta R\phi^2$, in this action. The equation of motion of this theory is given by,

\begin{align}
\begin{split} \label{emn}
(1+\zeta\, \phi^2)G_{\mu\nu} = \nabla_{\mu}\phi\nabla_{\nu}\phi-g_{\mu\nu}\Big(\frac{1}{2}(\nabla \phi)^2-\frac{12}{L^2}V(\phi)\Big)+\zeta\Big(g_{\m\nu}\Box \phi^2-\nabla_{\mu}\nabla_{\nu}\phi^2\Big) \\
\frac{1}{\sqrt{g}}\partial_{\mu}(\sqrt{g}g^{\mu\alpha}\partial_{\alpha}\phi)+\frac{12}{L^2}\frac{\partial V(\phi)}{\partial \phi}+\zeta\, \phi\, R =0
\end{split}
\end{align}

where $G_{\mu\nu}$ is the Einstein tensor. We can see from the equation of motion and also from the action that there is no clear separation between matter and gravity. In particular there is no unique way of defining matter stress tensor. For example one could define an effective matter stress tensor by solving for the Einstein tensor and defining, 

\be
[T_{\mu\nu}]_{eff} =  G_{\mu\nu} = \frac{1}{1+\zeta \phi^2} \Big[ \nabla_{\mu}\phi\nabla_{\nu}\phi-g_{\mu\nu}\Big(\frac{1}{2}(\nabla \phi)^2-\frac{12}{L^2}V(\phi)\Big)+\zeta\Big(g_{\m\nu}\Box \phi^2-\nabla_{\mu}\nabla_{\nu}\phi^2\Big)\Big]
\ee 

This effective matter stress tensor violates null energy condition for any non-zero value of the coupling parameter $\zeta$ \cite{Barcelo:2000zf}. In our framework this is not a problem as long as we have an entropy function for the gravity-matter system, described by (\ref{nmc}), which satisfies the second law. The validity of the second law does not depend on our ability to define a cleanly separated matter sector. 


Such an entropy function indeed exists for this theory which satisfies the second law to all orders. It is given by \cite{Ford},
\be \label{nonminimal}
S =\frac{2\pi}{\ell_{p}^3}\int d^3 x\sqrt{h} \Big(1-8\pi\,\zeta \phi^2\Big)
\ee


where the integral is over a space-like slice of the dynamical horizon. The form of the entropy function (\ref{nonminimal}) is valid for any form of the scalar self-interaction $V(\phi)$. Let us now assume that in the domain wall geometry the scalar field $\phi$ depends only on the coordinate $z$(or $\bar z$). Using this and the same steps as in the previous sections we arrive at the expression of the $c$ function to be,

\be \label{nonc}
a(\sigma)=\frac{2\pi\,L^{3}_{AdS}}{\ell_{p}^3}\Big(\frac{\bar z}{ z}\Big)^3\Big(1-8\pi \, \zeta \phi^2(z(\sigma))\Big)
\ee

It gives the correct central charge at the fixed point. In the above formula the scalar field $\phi$ is evaluated along the null generator of the causal horizon and so it is a function of the parameter $\sigma$. The monotonicity of the $c$-function (\ref{nonc}) follows from the second law \cite{Ford}. 

So the condition that a gravity-matter system satisfies the second law of causal horizon thermodynamics in AdS is more fundamental, because it is enough to prove the holographic $c$-theorem even in cases where null energy condition on matter stress tensor cannot be defined. Hence:

 \textit{The condition, which replaces the null energy condition on matter stress tensor, is that the second law of causal horizon thermodynamics \cite{Jacobson:2003wv} be satisfied by the gravity-matter system in AdS. This also includes theories with higher derivative couplings}. 
 
This condition has the virtue that this does not require a separation between the gravity and the matter sector because the validity of the second law does not require any such separation. We will now argue that this condition is sufficient for proving holographic $c$-theorem in an arbitrary gravity theory described by a local covariant action.

\subsection{Proof Of The Sufficiency Condition}\label{suffi}

In order to do that we consider an arbitrary local and diffeomorphism invariant theory in the bulk described by the action,

\be\label{hac}
I = \int d^{d+1}x \sqrt{-g} \ \mathcal{L} (g_{ab}, R_{abcd}, \nabla_e R_{abcd},..........,matter)
\ee
where various quantities have standard meaning. An action of this form will not in general allow a clean separation between matter and gravity sector and standard null-energy condition has no role to play.

Let us now assume that this theory admits several AdS critical points with constant values of the matter field which we assume to be scalar for simplicity. We consider a domain wall geometry of the form (\ref{domain}) which interpolates between a UV-AdS ($z\rightarrow 0$) and an IR-AdS ($z\rightarrow\infty$). This interpolating geometry represents the holographic RG flow \cite{Freedman:1999gp}. Now to prove the holographic $c$-theorem we consider causal horizons in this domain-wall geometry, following \cite{sb}. 

Let us assume that the theory of gravity described by (\ref{hac}) satisfies the second law of causal horizon thermodynamics. This means that there exists an entropy expression which satisfies the second law of thermodynamics for causal horizons and reduces to the \textbf{Wald Entropy} when the causal horizon is Killing. Wald entropy for the theory (\ref{hac}) is given by \cite{Wald,Iyer}, 

\be\label{wald}
S_W = -2\pi \int d^{d-1}y \sqrt\gamma \ \frac{\delta \mathcal{L}}{\delta R_{abcd}} \epsilon_{ab}\epsilon_{cd}
\ee 
where the integral is done over a codimension two space-like slice of the Killing horizon and $\gamma$ is the induced metric. $\epsilon_{ab}$ is the binormal to the horizon slice. We would like to emphasize that this is not the same entropy expression which satisfies the second law. We will use this formula later. 

 To proceed let us further assume that the dynamical entropy is again given by the integral of a local geometric expression on space-like slices of the dynamical (non-Killing) horizon. Schematically,

\be\label{dwald}
S_{dyn}(\sigma) = -2\pi \int_{[\sigma]} d^{d-1}y \sqrt{\gamma} \ s
\ee
where $\sigma$ is a real parameter, not necessarily affine, along the null generators of the horizon and $[\sigma]$ denotes the $\sigma=$ constant space-like slice. According to our assumption $s$ reduces to the Wald expression (\ref{wald}) when evaluated on a stationary horizon.  Since we are considering a dynamical horizon $S_{dyn}(\sigma)$ is a non-constant function of  $\sigma$ and the statement of the (local) second law is,

\be\label{dent}
\frac{d}{d\sigma} \Big(-2\pi \sqrt\gamma s \Big) \le 0
\ee  

We are getting a monotonic decrease rather than increase because we are looking at  a \textit{past} horizon. 

Let us now apply it to our case. We follow the same strategy as in \cite{sb} and in the previous sections, except that we are now doing it for an arbitrary theory (\ref{hac}) obeying the second law. We take the entropy expression (\ref{dent}) and evaluate it on the space-like cross-section (\ref{parahyp}) of the non-Killing causal horizon (\ref{dcausal}). $\sigma$ is the non-affine parameter, parametrizing the null generators (\ref{inte}) along which the space-like slice $[\sigma]$ is evolving. The induced metric $\gamma$ on $[\sigma]$ is given by (\ref{hyp}). Now in the domain wall geometry all the geometric quantities and the matter fields depend only on the $z$ (or $\bar z$) coordinate and so is constant on $[\sigma]$. As a result $s$ is also constant on $[\sigma]$ although it has non-trivial dependence on $\sigma$ through its dependence on $z$ (or $\bar z$). 


Now for the induced metric $\gamma$ we get from (\ref{hyp}),

\be
\sqrt\gamma \ d^{d-1}y = L_{AdS}^{d-1} \Big(\frac{\bar z}{z}\Big)^{d-1} dV_{H^{d-1}}
\ee

where $dV_{H^{d-1}}$ is the volume element of a unit hyperbolic space space. Since the intrinsic coordinates parameterizing the hyperbolic space (or $[\sigma]$) are comoving, $dV_{H^{d-1}}$ is independent of $\sigma$. So the second law ($\ref{dent}$) can be written as,

\be\label{monotony}
\frac{d}{d\sigma} \Big(-2\pi L_{AdS}^{d-1} \Big(\frac{\bar z}{z}\Big)^{d-1} s \Big) \le 0 
\ee

So we define our $c$-function to be, 

\be\label{a}
a(\sigma) = -2\pi L_{AdS}^{d-1} \Big(\frac{\bar z}{z}\Big)^{d-1} s(\sigma) 
\ee

The monotonicity of the $c$-function, $a'(\sigma)\le 0$, follows from ($\ref{monotony}$). Now we show that at the UV ($\sigma\rightarrow -\infty$) and the IR ($\sigma\rightarrow\infty$) fixed points the $c$-function ($\ref{a}$) gives the correct central charges \footnote{It follows from (\ref{inte}) that as $\sigma\rightarrow -\infty$, $\bar z\rightarrow 0$ and as $\sigma\rightarrow\infty$, $\bar z\rightarrow\infty$. It follows from a simple argument of \cite{sb} that $z\rightarrow 0$ as $\bar z\rightarrow 0$ and $z\rightarrow \infty$ as $\bar z\rightarrow \infty$. So as $\sigma$ runs from $-\infty$ to $+\infty$, the null generator of the causal horizon runs from the UV to the IR.}. Near the UV and the IR fixed points the causal horizon (\ref{dcausal}) again becomes \textbf{Killing} (stationary) because scaling is recovered. So to evaluate the central charge at the UV and the IR fixed points we can use the Wald entropy formula, i.e, 

\be
a (\sigma\rightarrow -\infty) = -2\pi L_{AdS}^{d-1} \Big(\frac{\bar z}{z}\Big)^{d-1} s(\sigma)\Big |_{\sigma\rightarrow -\infty} = -2\pi L_{AdS}^{d-1} \ s_{Wald}  \Big |_{UV-AdS}
\ee
and 
\be
a (\sigma\rightarrow \infty) = -2\pi L_{AdS}^{d-1} \Big(\frac{\bar z}{z}\Big)^{d-1} s(\sigma)\Big |_{\sigma\rightarrow\infty} =  -2\pi L_{AdS}^{d-1} \Big(\frac{L_{IR}}{L_{AdS}}\Big)^{d-1} \ s_{Wald}  \Big |_{IR-AdS}
\ee

where ,
\be
s_{Wald} = \frac{\delta \mathcal{L}}{\delta R_{abcd}} \epsilon_{ab}\epsilon_{cd}
\ee
and we have used the fact that $\frac{\bar z}{z}\rightarrow 1$ as $\sigma\rightarrow -\infty$ and $\frac{\bar z}{z}\rightarrow \frac{L_{IR}}{L_{AdS}}$ as $\sigma\rightarrow \infty$. The $s_{wald}$ has to be evaluated on the UV-AdS and the IR-AdS. This was evaluated in \cite{ctheorems} for a general bulk gravity theory described by the action (\ref{hac}). Their result is,

\be
\frac{\delta \mathcal{L}}{\delta R_{abcd}} \Big |_{AdS}= -\frac{\tilde L^2}{4d} \ (g^{ac}g^{bd} - g^{ad}g^{bc}) \ \mathcal{L}\Big |_{AdS} 
\ee
where $\tilde L$ is the AdS radius and everything is evaluated on pure AdS. Using this and the fact that $\epsilon_{ab}\epsilon^{ab} = - 2$, we get, 

\be
s_{Wald} \Big |_{AdS} = \frac{\delta \mathcal{L}}{\delta R_{abcd}} \epsilon_{ab}\epsilon_{cd} \Big |_{AdS} = \frac{\tilde L^2}{d} \ \mathcal{L} \Big |_{AdS}
\ee

Therefore, 
\be
a (\sigma\rightarrow -\infty) = -2\pi L_{AdS}^{d-1} \ s_{Wald}  \Big |_{UV-AdS} = -2\pi \frac{L_{AdS}^{d+1}}{d} \ \mathcal{L}\Big |_{UV-AdS}
\ee

and 

\be
a (\sigma\rightarrow \infty) = -2\pi L_{AdS}^{d-1} \Big(\frac{L_{IR}}{L_{AdS}}\Big)^{d-1} \ s_{Wald}  \Big |_{IR-AdS} =-2\pi \frac{L_{IR}^{d+1}}{d} \ \mathcal{L}\Big |_{IR-AdS}
 \ee

Due to maximal symmetry of pure AdS the Lagrangian density $\mathcal{L}$ evaluated on pure AdS is just a constant number. 

Now for even $d$ the coefficient of the Euler density \footnote{We are following the notation and conventions of \cite{ctheorems}.} in the trace-anomaly is given by \cite{Imbimbo:1999bj, ctheorems},

\be
A^{(d)} = - \frac{\pi^{\frac{d}{2}}}{\Gamma(\frac{d}{2})} \frac{\tilde L^{d+1}}{d} \ \mathcal{L}\Big |_{AdS}
\ee

So we can see that in even $d$, 

\be 
a (\sigma\rightarrow -\infty) = 2\pi \frac{\Gamma({\frac{d}{2}})}{\pi^{\frac{d}{2}}} \ A^{(d)}_{UV}
\ee

and 

\be 
a (\sigma\rightarrow \infty) = 2\pi \frac{\Gamma({\frac{d}{2}})}{\pi^{\frac{d}{2}}} \ A^{(d)}_{IR}
\ee

So these two equations together with the monotonicity condition ($\ref{monotony}$) $a'(\sigma)\le 0$ following from the second law, proves the holographic $c$-theorem for even $d$. 

For odd $d$ there is no trace-anomaly. However, as shown in \cite{ctheorems,Myers:2010xs}, for odd $d$, the number $A^{(d)}$ can be interpreted in terms of the entanglement entropy of the dual conformal field theory. We refer to these references for further details.

\section{Higher derivative theories and second law for causal horizons in AdS}


The calculations in \cite{sb} and in the previous sections show that whenever the second law of causal horizon thermodynamics holds in a theory of gravity coupled to matter in $AdS$, the entropy density evaluated on the deformed (non-Killing) causal horizon ($\ref{dcausal}$) is a holographic $c$-function. This is consistent with the proposal of \cite{sb} which says that the boundary RG flow is dual to the thermodynamics of causal horizons, i.e, the validity of the second law is both a \textbf{necessary} and a \textbf{sufficient} condition for holographic $c$-theorem to hold. We have proved the sufficiency condition in the last section for a general theory of gravity in the bulk. This raises an interesting puzzle. If the dual of the boundary $c$-theorem is the second law of thermodynamics satisfied by causal horizons in $AdS$, then what happens to bulk theories which violate the second law ? If the bulk theory is $\textit{non-unitary}$ then this is not a problem, because the dual boundary theory is also non-unitary and $c$-theorem does not necessarily hold. But if the bulk theory is unitary then the dual field theory is unitary and $c$-theorem must hold . So :

 \textit{If the duality between $c$-theorem and the second law \cite{sb} is correct, then this implies that every \textbf{unitary} bulk theory of gravity coupled to matter, \textbf{including those with higher-derivative couplings}, satisfies the \textbf{second law of causal horizon thermodynamics} in $AdS$. If this is not true then $c$-theorem will be violated in a unitary Lorentz invariant field theory}. \footnote{The first law for AdS-Rindler horizon follows from the positivity of the Relative entropy in the dual field theory as shown in \cite{Lashkari:2013koa,Faulkner:2013ica, Bhattacharya:2013bna}. }
 
 In any case, we hope to have convinced the reader that there is a deep connection between second law of causal horizon thermodynamics and unitarity and Lorentz invariance of the holographic dual field theory. This needs to be fleshed out.

 Moreover our proof of the sufficiency condition in (\ref{suffi}) shows that for a unitary bulk theory the \textit{density} of the dynamical entropy as defined in (\ref{dwald}) must be positive, i.e, $-s\ge 0$. If this condition is violated at any point of the dynamical causal horizon then the holographic $c$-function as defined in (\ref{a}) will be negative. This negativity is a signature of non-unitarity. The monotonicity of the $c$-function also requires the positivity of the entropy production at every point of the causal horizon. This is usually satisfied in all known cases but this could be a non-trivial constraint on a general theory of gravity in the bulk. Hence \textit{\textbf{unitarity} and Lorentz invariance of the holographic dual field theory requires a positive entropy density which is non-decreasing at every point of the dynamical causal horizon}.

  Non-unitary theories can also satisfy the second law, but it is not a necessary consequence of the duality between $c$-theorem and the second law. Let us now say a few words about the implication of this for higher-derivative theories of gravity in $AdS$. 
 
 \begin{figure}[htbp]
\begin{center}
  \includegraphics[width=13cm]{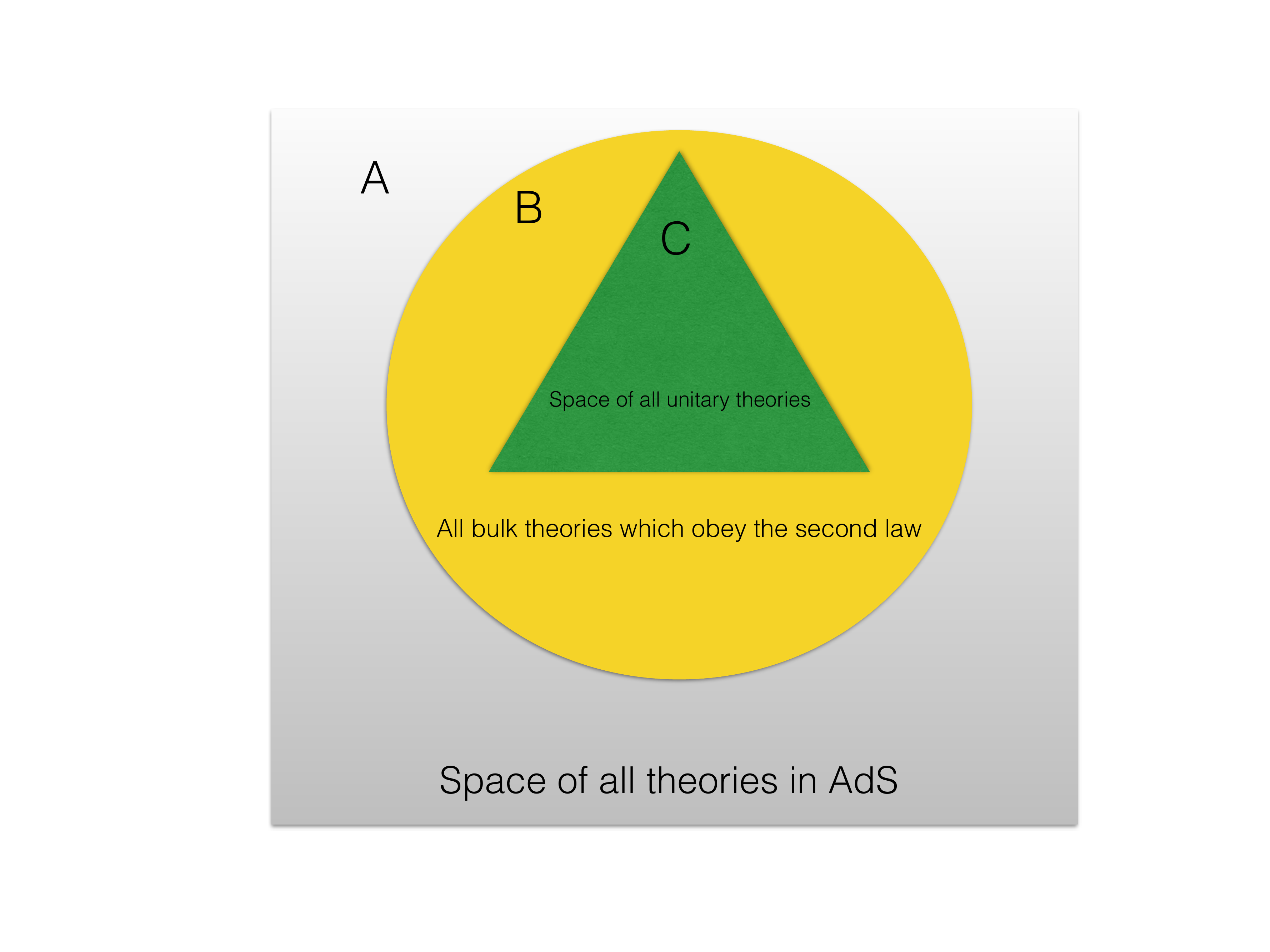}
\end{center}
\caption{Square $A$ = Space of all theories in AdS ; $A \supseteq B$ = Space of all theories which obey the second law of causal horizon thermodynamics in AdS ; $B\supseteq C$ = Space of all \textbf{unitary} theories in AdS . }
\end{figure}
 
 In higher derivative theories the entropy of a  (bifurcate Killing horizon) stationary black holes is given by the Wald entropy \cite{Wald, Jacobson:1993xs, Jacobson:1993vj, Iyer} which by construction satisfies the first law of thermodynamics. The story of the second law is different. It is known that the Wald entropy does not in general satisfy the second law.  There are some cases \cite{jkm,Sarkar,Sarkar1,Sarkar2} where an entropy function exists which satisfies the second law and reduces to the Wald entropy in the stationary situation, but no general expression exists. In fact it is an open question if every higher derivative theory of gravity satisfies the second law or not. We have argued that for causal horizons in $AdS$ the second law must be satisfied by every \textit{unitary} higher-derivative theory of gravity as a consequence of the duality between $c$-theorem and the second law. This is an indirect holographic argument and uses unitarity crucially. Figure-3 summarizes our claim.

An interesting question is whether $A=B$ or $B$ is a proper subset of $A$. That is whether every classical gravity theory satisfies the second law of causal horizon thermodynamics or not ? By every classical gravity theory we mean every covariant unitary or non-unitary gravity plus matter theory in AdS described by an action of the form (\ref{hac}). We do not know if this is true or not, but our results show that if this is true, then : 

\textit{Every unitary or non-unitary field theory with a classical gravity dual satisfies the $c$-theorem. In other words, in the large-N but finite $\lambda$ limit, $c$-theorem becomes insensitive to the unitarity property of field theory}. Here $\lambda$ is the 't Hooft coupling of the dual field theory.


\section{discussion}

Before we conclude we would like to discuss some possible extension of our work. One such thing is to study the $c$-function in a thermal state which is dual to a black hole geometry \cite{Paulos:2011zu}. Another potential application is to the study RG flows in the attractor geometry \cite{Goldstein:2005rr}. Another important problem is to understand the proper field theory interpretation of the results of \cite{sb} and those presented in this paper. In short, our results show that at least in the large-$N$ limit RG flow and $c$-theorem has an emergent thermodynamics interpretation where the $c$-function plays the role of the entropy density. In fact the $c$-function at the fixed point has a similar interpretation found in \cite{ctheorems,Myers:2010xs,chm}. Our results show that such an interpretation must exist even away from the fixed point \footnote{SB would like to thank Kristan Jensen for useful discussion on this.}. It will also be interesting to see how boundary RG flow \cite{Affleck:1991tk} fits into this framework. We hope to return to them in near future.

\section{Acknowledgements}
We thank Aninda Sinha for useful discussions and collaboration in the early stages of this work. AB would like to thank Sudipta Sarkar and Srijit Bhattacharjee for illuminating discussions. AB thanks IIT-Gandhinagar for hospitality during the early stages of this work. AB also like to thank Institute of Physics, Bhubaneswar where the part of this work is presented in the \textit{Young Researchers' Conference '15}. SB would like to thank Jyotirmoy Bhattacharya, Simeon Hellerman, Kentaro Hori, Raghu Mahajan, Djordje Radicevic, Steve Shenker and Masahito Yamazaki for very helpful discussions and insightful questions. SB would also like to thank the string theory group at Stanford University (SITP) for hospitality during the initial stages of this project. SB would like to thank Shiraz Minwalla for an interesting discussion on second law in higher derivative gravity during a visit to TIFR, Mumbai. We would also like to thank Swagata Ghatak for her assistance with the figures. The work of SB was supported by World Premier International Research Center Initiative (WPI), MEXT, Japan.

\appendix
\section{Monotonicity of $c$ function for the general curvature squared theory }
In this section we will demonstrate the monotonicity  of (\ref{c2}) for a certain restricted class of curvature squared theories following \cite{ctheorems}. The constraint can be imposed by setting $a_1=0$.

 We first write down the equation of motion for this theory. 
\be \label{met3a}
R_{\alpha\beta}-\frac{1}{2} R g_{\alpha\beta}-\frac{6}{L^2} g_{\alpha\beta}-\frac{L^2}{2}\, H_{\alpha\beta}=T_{\alpha\beta}
\ee
where,
\begin{align}
\begin{split}
H_{\alpha\beta}= &\,\,
\lambda_{3}\Big(2\nabla_{\alpha}\nabla_{\beta} R-2 RR_{\alpha\beta}+\frac{1}{2}g_{\alpha\beta}(R^2-4\nabla_{\gamma}\nabla^{\gamma} R)\Big)+\\&\lambda_{2}\Big(\nabla_{\alpha}\nabla_{\beta}R+2R_{\gamma\alpha\beta\delta}R^{\gamma\delta}-\nabla^2 R_{\alpha\beta}+\frac{1}{2}g_{\alpha\beta}(R_{\gamma\delta}R^{\gamma\delta}-\nabla^2 R)\Big)+\\&\lambda_{3}\Big(2R_{\alpha\gamma\delta\zeta}R^{\gamma\delta\zeta}{}_{\beta}+2\nabla_{\alpha}\nabla_{\beta}R+4 R_{\alpha\gamma}R^{\gamma}{}_{\beta}-4 R_{\alpha\gamma\beta\delta}R^{\gamma\delta}-4\nabla^2 R_{\alpha\beta}+\frac{1}{2}g_{\alpha\beta}R_{\gamma\delta\zeta\eta}R^{\gamma\delta\zeta\eta}\Big)\n.
\end{split}
\end{align}
 and $T_{\alpha\beta}$ is the matter stress tensor.  We have assumed that the matter stress-tensor obeys the null-energy condition. So if we contract this with the null vector $\bar D$ we get,
\be \label{met2a}
R_{\alpha\beta}\bar D^{\alpha}\bar D^{\beta}-\frac{L^2}{2}\,H_{\alpha\beta}\bar D^{\alpha}\bar D^{\beta}\geq 0.
\ee
Evaluating (\ref{met2a}) on (\ref{remetric}) we get the following condition after evaluating on the horiozn $-t^2+r^2+\bar z^2=0$,
\begin{align}
\begin{split} \label{met4}
&\Big(\frac{z''\bar z^2 \left(6+ f_{\infty} (4 \lambda_{1}+2\lambda_{2}-8 \lambda_{3}+a_{1}) z'^2\right)}{8 z}+\frac{1}{8} z^{(3)} f_{\infty} \bar z^2  z' a_{1}\\&-\frac{1}{8} z^{(4)} f_{\infty} \bar z^2 z a_{1}+\frac{7}{8} z''^2 f_{\infty} \bar z^2 a_{1}\Big)\geq\,0.
\end{split}
\end{align}
where, $a_{1}=(4 \lambda_{1}+5 \lambda_{2}+16 \lambda_{3}).$   When  $a_{1}=0$ the radial equation of motion simplifies and all the terms containing more than two derivatives of $z$ cancel \cite{ctheorems}.  So setting $a_1=0$ we get,
\begin{align}
\begin{split} \label{met4a}
\frac{6 z'' \bar z^2 \left(1-\frac{1}{2} f_{\infty} (\lambda_{2}+8 \lambda_{3}) z'^2\right)}{8 z}\geq\,0.
\end{split}
\end{align}
Then applying the lineariztion to (\ref{met4a}) we get,
\be
f'(z)(1-\frac{1}{2}f_{\infty}(\lambda_{2}+8\lambda_{3})f(z))\geq\,0.
\ee

\par

Next we evaluate,
\begin{align}
\begin{split} \label{er}
\frac{d a(\sigma)}{d\sigma}=&\,\, \frac{3 \bar z \left(z-\bar z z'\right) \left(\bar z^2 \left(4-5 f_{\infty} a_{1} z'^2\right)+f_{\infty} (4 \lambda_{1}+3 \lambda_{2}) z^2\right)}{4 z^4}\\&+\frac{z'' a_{1}\left(f_{\infty} \bar z^3 \left(3 z-7 \bar z z'\right)\right)}{2 z^3}+\frac{z^{(3)}(\bar z) f_{\infty} \bar z^4 a_{1}}{2 z^2}
\end{split}
\end{align}
Again setting $a_{1}=0$ we get,
\begin{align}
\begin{split} \label{er1}
\frac{d a(\sigma)}{d\sigma}=&\,\, 3 \frac{\bar z^3}{z^3} \left(1-\frac{\bar z}{z} z'\right) \left( 1 -\frac{1}{2} f_{\infty} (\lambda_{2}+8 \lambda_{3}) \left(\frac{z}{\bar z}\right)^2\right)
\end{split}
\end{align}
Linearization gives,
\begin{align}
\begin{split} \label{er2}
\frac{d a(\sigma)}{d\sigma}=&\,\, 3  \left(1-\int_{0}^{1}\frac{\sqrt{f(z)}}{\sqrt{f(\alpha z)}} d\alpha \right) \left( 1-\frac{1}{2} f_{\infty} (\lambda_{2}+8 \lambda_{3}) \right)
\end{split}
\end{align}
Now we can write,

\be
1-\int_{0}^{1}\frac{d\alpha\, \sqrt{f(z)}}{\sqrt{f(\alpha z)}}  = - \frac{1}{2} \int_{0}^{1} \frac{d\alpha}{\sqrt{f(\alpha z)}} \int_{\alpha z}^{z} \frac{d\beta}{\sqrt{f(\beta)}} f'(\beta) 
\ee

Using this we can write, 

\be \label{Savariation1}
\frac{d a(\sigma)}{d\sigma}= -\frac{3}{2} \int_{0}^{1} \frac{d\alpha}{\sqrt{f(\alpha z)}} \int_{\alpha z}^{z} \frac{d\beta}{\sqrt{f(\beta)}} f'(\beta) 
 \left( 1-\frac{1}{2} f_{\infty} (\lambda_{2}+8 \lambda_{3}) \right).
\ee
So,
\be
\frac{d a(\sigma)}{d\sigma}\leq\,0
\ee
upto the linear order.

\end{document}